# THGEM-based detectors for sampling elements in DHCAL: laboratory and beam evaluation

**L. Arazi**[a,*], **A. Breskin**[a], **R. Chechik**[a], **M. Cortesi**[a,b], **M. Pitt**[a], **A. Rubin**[a]

[a] *Weizmann Institute of Science, P.O. Box 26 Rehovot 76100, Israel*
[b] *Present address: Paul Scherrer Institut, 5232 Villigen PSI, Switzerland*

**H. Natal da Luz**[c], **J. M. F. dos Santos**[c]

[c] *University of Coimbra, Portugal*

**C. D. R. Azevedo**[d], **D. S. Covita**[d], **C. A. B. Oliveira**[d], **J. F. C. A. Veloso**[d]

[d] *I3N, Department of Physics, University of Aveiro, Portugal*

**M. Breidenbach**[e], **D. Freytag**[e], **G. Haller**[e], **R. Herbst**[e]

[e] *SLAC, Stanford, USA*

**S. Park**[f], **A. White**[f], **J. Yu**[f]

[f] *University of Texas at Arlington, USA*

**E. Oliveri**[g]

[g] *University of Siena- INFN Pisa, Italy*

[*]*Corresponding author*
*E-mail*: `lior.arazi@weizmann.ac.il`

ABSTRACT: We report on the results of an extensive R&D program aimed at the evaluation of Thick-Gas Electron Multipliers (THGEM) as potential active elements for Digital Hadron Calorimetry (DHCAL). Results are presented on efficiency, pad multiplicity and discharge probability of a 10x10 cm$^2$ prototype detector with 1 cm$^2$ readout pads. The detector is comprised of single- or double-THGEM multipliers coupled to the pad electrode either directly or via a resistive anode. Investigations employing standard discrete electronics and the KPiX readout system have been carried out both under laboratory conditions and with muons and pions at the CERN RD51 test beam. For detectors having a charge-induction gap, it has been shown that even a ~6 mm thick single-THGEM detector reached detection efficiencies above 95%, with pad-hit multiplicity of 1.1-1.2 per event; discharge probabilities were of the order of $10^{-6}$-$10^{-5}$ sparks/trigger, depending on the detector structure and gain. Preliminary beam tests with a WELL hole-structure, closed by a resistive anode, yielded discharge probabilities of <$2\times10^{-6}$ for an efficiency of ~95%. Methods are presented to reduce charge-spread and pad multiplicity with resistive anodes. The new method showed good prospects for further evaluation of very thin THGEM-based detectors as potential active elements for DHCAL, with competitive performances, simplicity and robustness. Further developments are in course.

## Contents

## 1. Introduction

Digital hadron calorimetry (DHCAL) has been proposed for facilitating and improving jet-energy resolution in precise physics measurements planned for future high energy colliders, such as the International Linear Collider (ILC) [1,2]. Digital recording of jet-induced hits with pixelated gas sampling elements, accompanied by advanced pattern recognition algorithms are expected to yield very high-precision jet-energy measurements as simulated by CALICE: $\sigma/E_{jet}$ ~3-4% [1,2].

    The baseline design of the hadronic calorimeter of the ILC Silicon Detector (SiD) comprises 40 layers of stainless steel absorber plates separated by 8 mm gaps, which should incorporate the active detection elements and their readout electronics [1]. The target jet-energy resolution calls for high detection efficiency and low average pad multiplicity (number of pads triggered per particle). Active elements utilizing Resistive Plate Chambers (RPCs), the baseline technology for the SiD hadronic calorimeter, have yielded so-far an average multiplicity of 1.5-2 at 90-95% efficiency [3]. Detection elements based on the MICRO MEsh Gaseous Structure (MICROMEGAS), have demonstrated 98% efficiency with a 1.1 average multiplicity [4]. Elements based on double Gaseous Electron Multipliers (double-GEM), have shown a multiplicity of ~1.3 at 95% efficiency [5].

    In this work we consider the possible use of Thick-GEMs (THGEMs) [6, 7] as the basic building blocks for DHCAL sampling elements. THGEMs are particularly attractive because of their simplicity and robustness, accompanied by sub-mm spatial resolution, few-ns temporal resolution and up to 1 MHz/mm$^2$ rate capability [8, 9]. In what follows, we relate to experience gained in recent beam tests, primarily with thin, single-THGEM, and with thicker, double-THGEM elements, followed by metal pads coupled to the KPiX [10] highly-integrated analog readout electronics. Some investigations involved a very thin configuration, WELL, of a single-THGEM, having its bottom side in direct contact with the readout pads. Results of preliminary beam tests and laboratory R&D work done on structures comprising resistive anodes, conceived for discharge damping, are briefly summarized.

## 2. Experimental setup and methodology

### 2.1 Beam tests

Beam tests were conducted at the CERN SPS/H4 RD51 beam-line, with different THGEM-based detector configurations. The THGEM electrodes used in this work were 10x10 cm$^2$ in size, manufactured by 0.5 mm diameter hole-drilling in 0.4 mm thick G10 plates, Cu-clad on one or two sides; the holes were arranged in an hexagonal lattice with a pitch of 1 mm; 0.1 mm wide rims were chemically etched around the holes. These parameters were chosen based on previous optimization studies [7]; in view of previous experience with neon mixtures [9, 11], the detectors were operated in Ne/5%CH$_4$, where minimally ionizing particles (MIPs) in the relevant energy range deposit on the average a total number of ~60 electrons per cm along their track [12]. The experiments were performed with 150 GeV/c muons and pions.

The main detector configurations investigated comprised either a single-THGEM, (with a drift gap of 3-4 mm and an induction gap of 2 mm), or a double-THGEM (with a 3 or 10 mm drift gap, 2 mm transfer gap and 2 mm induction gap) (fig. 1). Single-THGEM operation at beam conditions was done at a gain of 1-2×10$^3$; the double-THGEM detector was operated at a total gain of 4-8×10$^3$. The respective voltage settings are provided in the results section.

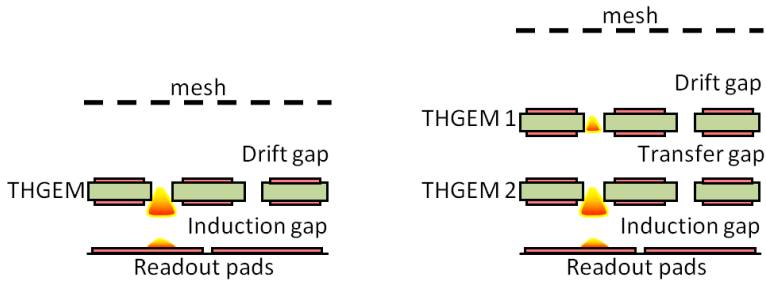

**Figure 1:** Single- and double-THGEM configurations. The metal readout pads collect the direct avalanche charge.

The detector was operated with an external trigger, provided by the RD51 GEM/MICROMEGAS tracker telescope setup [13, 14] with its three 10x10 cm$^2$ scintillators in coincidence. In some measurements, a fourth 0.5x1 cm$^2$ scintillator, set in front of the detector's central region was added to the coincidence; it assured triggering on particles passing through a small active area of the detector, eliminating "empty" triggers with recorded electronic noise. The HV power supply was remotely controlled with the CAEN SY2527 controller, using A1833P and A1821N boards. The voltage and current on each channel were monitored using dedicated RD51 slow control system [15]. All inputs were connected through RC filters (a 15 MΩ resistor + 1 nF capacitor).

Initial beam studies aimed at characterizing the intrinsic physical behavior of the detector in beam conditions, decoupled from the particular choice of a readout system. These included gain, efficiency, pulse-height distributions and discharge-probability measurements, using discrete electronics for signal readout and processing. The anode consisted of four central 1 cm square Cu pads, interconnected to a charge sensitive preamplifier, linear shaping amplifier and multichannel analyzer (MCA). In another configuration, the anode comprised 16 pads, each connected to a separate charge sensitive preamplifier (Cremat CR-110 with CR-150 evaluation board) followed by a multi-channel digitizer (VME 3212/3213), as described in [16].

The core of the beam studies was performed using an array of 8×8 1 cm square Cu pads, coupled to the KPiX readout electronics [10]. KPiX is a multi-channel system-on-chip, for self-triggered detection and processing of low-level charge signals, whose development was motivated by the SiD for the ILC. Data acquisition proceeds autonomously under control of a digital core on receipt of a signal initiating an acquisition cycle. A variety of operating conditions can be selected by preloading a control register. (Many features important for ILC operation are not covered in this short summary). Because the intended application is for a pulsed accelerator, data acquisition proceeds for a period of up to 3 ms, followed by readout of digital data. This resulted in a limited live-time of ~3% for the present test runs. Amplitude and timing information for up to four events per acquisition cycle can be collected for each pixel. The version of KPiX used in these tests contained 512 channels, of which only 64 were wire-bonded to the 8 x 8 array of pads[1]. Because of the quasi-continuous nature of the test beam, the DC-reset option was used; acquisition was done with external trigger using the RD51 tracker telescope. For KPiX bonded to the pad array inside the detector chamber, the measured noise level was ~0.3 fC. This included contributions from the external protection elements, resistors and diodes, which were installed to divert spark energy from the chip. The noise was Gaussian over five orders of magnitude. Noise signals collected for empty triggers (see below) could be removed by making a cut at ~1.5 fC. The gain of the THGEM detector thus needed to be large enough to shift the peak of the Landau distribution considerably beyond 1.5 fC.

Studying the possible damping of occasional spurious discharges, a preliminary experiment was also done in beam conditions with a resistive anode in contact with the THGEM, in a configuration named Resistive WELL, or RWELL for short (fig. 2). The avalanche development in this geometry is similar to that described in [17-19]. The RWELL configuration comprised one single-faced THGEM (with Cu on its top side only), with the bottom side in contact with a resistive layer (graphite particles mixed with an epoxy binder with a surface resistivity of ∼10 MΩ/square), deposited on a 0.2 mm thick G10 sheet. The sheet was set above the readout pad array which, in this case, picked up avalanche-induced charge through the resistive film, rather than the direct charge. The THGEM parameters were similar to those described above.

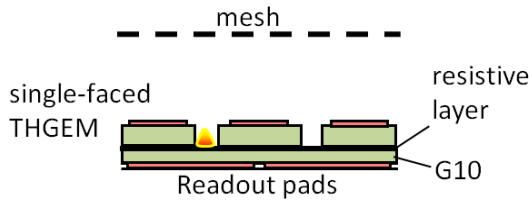

**Figure 2**: Resistive WELL (RWELL) configuration. The single-faced THGEM is set on top of a resistive layer (here 10 MΩ/square) deposited on a thin G10 sheet (here 0.2 mm thick), laid on top of the readout pads. The pads pick up an induced signal, rather than directly collecting the avalanche charge.

The data collected at the beam included pulse-height distributions, detection efficiency curves (with respect to the trigger), discharge-probability estimates and pad multiplicity. Studies of pad response versus event location were planned for assessing local efficiency and multiplicity, in particular for particles passing near the pad boundaries; these required event-matching between KPiX electronics and the tracker. A clock signal generated by the KPiX FPGA was sent to both KPiX and the tracker; thus, each KPiX event was assigned a time stamp, with a corresponding one given the tracker event. Data analysis of these measurements is in process; therefore, the data presented in this work were analyzed differently, as discussed below.

---

[1] A 1024-channel version of KPiX is available and preparations are under way to instrument a GEM pad-array of that size.

All experiments described in this work were performed with Ne/5%$CH_4$, at gas-flow rates of a few l/h.

## 2.2 Laboratory R&D on resistive-anode structures

The underlying motivation for using a resistive film as an intermediate layer between the THGEM and readout pads [20, 21] is twofold: (1) decoupling of the sensitive readout electronics from direct currents during occasional energetic discharges, and (2) the ability of the resistive layers to effectively reduce the spark energy, as demonstrated in previous works [22, 23]. In this article, we briefly describe the methodology and main results, referring the reader to a more complete work [24].

Laboratory studies on structures comprising a resistive anode focused on 3x3 $cm^2$ THGEM electrodes with variants of the basic RWELL structure described above (fig. 2). The readout electrode comprised four 1 $cm^2$ Cu pads. Initial investigations were carried out with 1-10 MΩ/square resistive layers deposited on a 0.2 mm thick G10 sheet. In further experiments the resistive layer was deposited on a grid of 200 μm wide Cu lines, printed on the thin G10 sheet, with the grid matching the boundaries between readout pads located behind the sheet (gridded-RWELL, or GRWELL); its role was to allow for rapid draining of avalanche electrons reaching the resistive layer at these places, thus reducing their spread onto neighboring pads [24]. This resistive-anode structure was initially coupled to a regular single-faced THGEM in WELL configuration with an hexagonal lattice of holes covering its entire active area; it was subsequently investigated with a segmented-THGEM electrode having a square lattice of holes covering only the pads area, with ~1 mm wide metal bands above the underlying resistive-anode grid lines (fig. 3). This configuration (segmented-GRWELL) intended avoiding avalanche occurrence above the grid lines, where the rapid clearance of avalanche electrons was shown to result in significantly reduced amplitude of the signal induced through the (locally lower-resistivity) film on the readout pads; the mismatch between holes and these strips also reduced the occurrence of eventual non-damped discharges [24].

The experimental work done with resistive-anode structures comprised an extensive series of measurements, of which we discuss here the gain, discharge probability, counting-rate capability, cross-talk between pads and signal amplitude as a function of the distance from the pad boundary. The bulk of the experiments was done with a collimated beam of 8 keV x-rays generated by an x-ray tube with a Cu target; signals from the 1 $cm^2$ readout pads were processed with a charge sensitive preamplifier, followed by a linear amplifier and an MCA or a digital oscilloscope. As a reference, measurements were also performed on a bare WELL, where a single-faced THGEM (with the same thickness and hole pattern as those of the resistive WELL), was set directly on top of the metallic readout pads. All measurements were done in Ne/5%$CH_4$ at 1 atm. Position-dependent charge measurements were performed by scanning the collimated (0.5 mm diameter) x-ray beam across the pads.

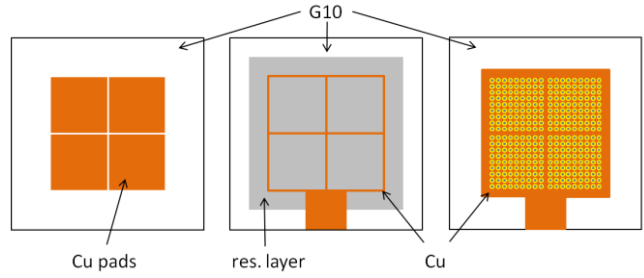

**Figure 3:** A segmented-GRWELL layout: Left – the Cu pad array (connections not shown). Middle - the gridded resistive anode: the resistive layer is deposited on top of a grid of copper lines printed on a thin G10 sheet. The grid lines, matching the pads' boundaries, rapidly drain avalanche electrons reaching the resistive layer, to reduce their diffusive spread to neighboring pads. Right – the segmented single-faced THGEM, with a square-hole pattern; the hole-less zones between the active THGEM ones, matching the grid lines of the resistive anode, prevent avalanche formation in their vicinity (see text).

## 3. Results

### 3.1 Beam tests

Prior to our recent investigations with KPiX readout, the 10×10 cm$^2$ detector properties in different configurations were studied with discrete electronics. A complete discussion of the test-beam results obtained with single- and double-THGEM configurations with discrete electronics is given in a separate publication [16].

Figure 4a shows an example of the pulse-height distribution recorded with muons in a narrow single-THGEM configuration (3 mm drift gap, 2 mm induction gap). Here, the charge signals from the central four 1 cm$^2$ pads interconnected were recorded with a single charge sensitive preamplifier/linear-amplifier/MCA. Data acquisition was triggered by the three telescope scintillators, in coincidence with a 0.5×1 cm$^2$ scintillator positioned in front of the detector's center. The detector was operated at a gain of ~1.1×10$^3$, with negligible discharge probability (of the order of 10$^{-6}$ or lower). The detection efficiency (recorded events per number of triggers) as a function of THGEM voltage is shown in figure 4b. Note that in this configuration, the total width of the detector was 5.5 mm.

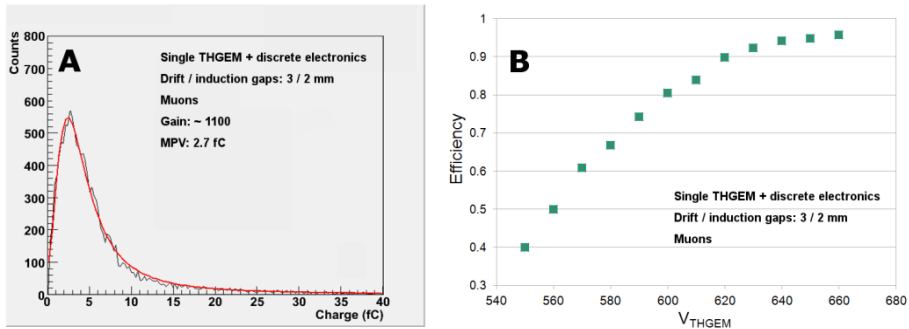

**Figure 4:** (A) Charge distribution of muons recorded with a 10×10 cm$^2$ single-THGEM detector (3 mm drift and 2 mm induction gaps) using a charge sensitive preamplifier connected to 4 pads. Ne/5%CH$_4$; gain ~1.1×10$^3$. The line matching the distribution is a Landau fit to the data. (B) Muon detection efficiency vs. THGEM voltage for the same configuration.

Initial investigations with KPiX were done with double-THGEM configurations, which allowed operation at relatively high gains with low discharge rates. Figure 5 shows examples of Landau distributions recorded with KPiX in double-THGEM structures, with a broad muon beam covering the entire detector area and with a narrow pion beam of a few mm in diameter impinging on the central area. The respective local rates for muons and pions were a few Hz/cm$^2$ and a few kHz/cm$^2$. The detector was investigated with drift gaps of 3 and 10 mm; in both cases the transfer and induction gaps were 2 mm wide. The applied voltages were asymmetric, with 610 V across the first THGEM and 510 V on the second. The drift-field was 650 V/cm for the 10 mm drift gap and 200 V/cm for the 3 mm one. In this configuration and voltages, the effective gain was in the range $4\text{-}8\times10^3$. The pulse-height distributions were obtained by recording, for each trigger, the sum of charge over the pad which displayed the highest signal and its 4 neighbors. The data shown in fig. 5 is preliminary, in the sense that it contains empty-trigger events with particles passing outside of the detector's active area. In this case, KPiX simply stored the amplitude of the noisiest channel, shown as the noise peak in the lowest charge bins. In principle, this effect can be mitigated by discarding events whose locations in the tracker fall outside of the region overlapping the detector's active area. This line of analysis is still in process.

All of the double-THGEM configurations displayed low discharge rates, with probabilities $< 2\times10^{-6}$ sparks/event - for both muons and pions (for both 3 and 10 mm drift gaps), as averaged over 7-8 hours of operation in beam per given configuration. Since in some runs there were no discharges, this value can be considered as an upper limit on the actual discharge probability for the double-THGEM structures in these operation conditions.

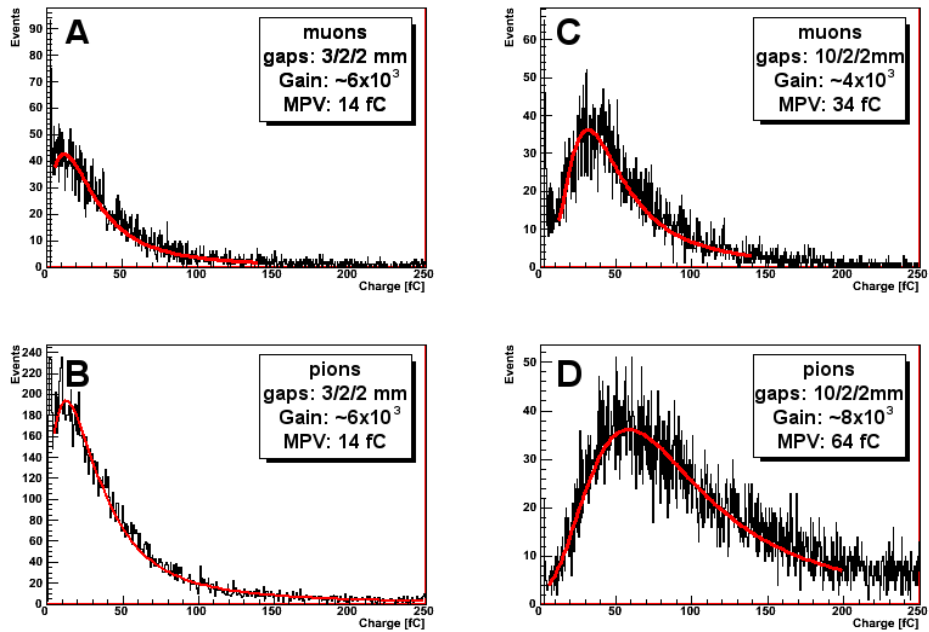

**Figure 5:** Landau distributions of muons and pions, recorded with KPiX, in two double-THGEM configurations with 3 mm (A,B) and 10 mm (C,D) drift-gap. Detector effective gains and most probable values (MPVs) are indicated in the figures. The muon rates (broad beam) were of a few Hz/cm$^2$; that of the pions (narrow beam) were a few kHz/cm$^2$. (See text for more details).

Further investigations were performed with a single-THGEM configuration of a 4 mm drift gap and 2 mm induction gap. To resolve the most probable value (MPV) from the empty-trigger noise, the detector was operated at a gain of ~$3\times10^3$, resulting in higher discharge probabilities: ~$1\times10^{-5}$ for muons and ~$1\times10^{-4}$ for pions. Figure 6a shows the Landau spectrum obtained in this configuration for a broad muon beam and figure 6b – for a narrow pion beam. In both cases the THGEM voltage was 700V, the drift field was 325 V/cm and the induction field 1 kV/cm. KPiX triggers following a discharge were removed from the analysis. The empty-trigger noise was removed in this case by placing a cut at 1.5 fC and the Landau fit was forced to go through zero, yielding an MPV of 9 fC for both muons and pions.

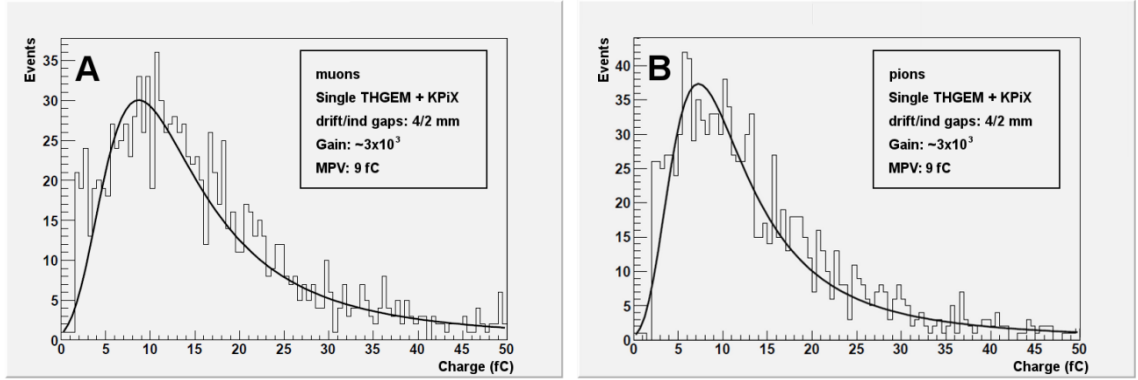

**Figure 6:** Landau distributions recorded with KPiX, in a single-THGEM configuration with a 4 mm drift gap and 2 mm induction gap: (A) broad muon beam; (B) narrow pion beam. Empty triggers were removed by setting a cut at 1.5 fC. Discharge probabilities were ~$1\times10^{-5}$ for muons and ~$1\times10^{-4}$ for pions at detector gain of $3\times10^3$. (See text for more details).

The average pad multiplicity (average number of pixels triggered per event) was estimated for the single-THGEM runs, by counting the number of neighboring pixels with a signal above 1 fC for all events where the maximum signal (main hit) was above 1.5 fC. Multiple-pixel hits during sparks were removed from the analysis. The resulting average pad multiplicity was 1.14 ± 0.03 for muons and 1.18 ± 0.02 for pions.

Efficiency measurements with KPiX in external trigger mode require either event matching between the detector and tracker, or the addition of a small scintillator in coincidence with the tracker telescope, to enable discarding empty triggers. For the present beam tests, work is still in process on event matching between the two systems. However, since it has already been demonstrated that a single-THGEM with a 3 mm drift gap can reach 96% efficiency (fig. 4), it can be safely estimated that a similar level of efficiency can be obtained for a single-THGEM detector coupled to KPiX with the empty triggers removed, since the typical KPiX noise is ~0.3 fC. Because of this low noise, it should be possible to operate the single-THGEM detector at a low gain, with low discharge probabilities, similar to those observed with discrete electronics. One should recognize, however, that multiplicity and efficiency are coupled through the threshold, and hence setting a low threshold to obtain a high efficiency may lead to higher multiplicity values. To get the multiplicity level mentioned above (1.1-1.2), the threshold should be set to ~1-1.5 fC. By calculating the area below the Landau distribution function, one can show that this would require the MPV to be at ~3-5 fC (detector gain of ~1-2×10$^3$) to keep the efficiency at ~95% or higher.

As noted above a preliminary test was also done with the RWELL configuration in beam conditions. Figure 7 shows the Landau distribution obtained with an RWELL (fig. 2), with the discrete readout electronics described above; all four pads were interconnected to the preamplifier. Unlike the regular single-THGEM structure, the WELL with resistive anode allowed raising the gain to ~4×10$^3$ (THGEM voltage of 710 V) with no evidence of discharges under a muon beam (discharge probability < 2×10$^{-6}$). Increasing the THGEM voltage to 720 V resulted in the onset of low-energy discharges, inducing typical current pulses of a few dozen nA (one order of magnitude lower than that observed in the regular single-THGEM configuration). The typical discharge-induced voltage drop was negligible (< 1 V) – two orders of magnitude lower than in the regular single-THGEM. The operation of an RWELL with KPiX is still under laboratory studies and was not yet tested in a beam.

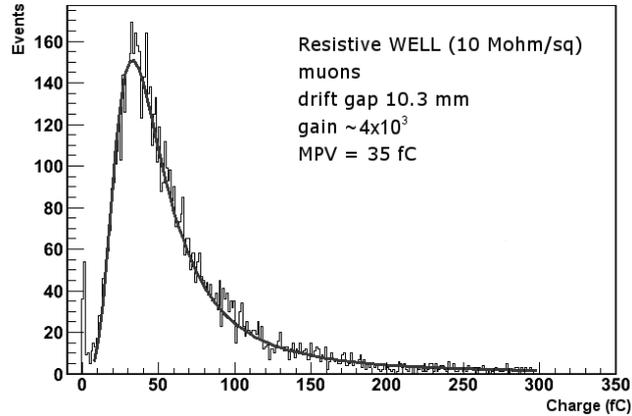

**Figure 7:** Landau distribution recorded with the RWELL configuration of Fig. 2. THGEM voltage was 710 V. No discharges were observed during the measurement (discharge probability < 2×10$^{-6}$).

### 3.2 Laboratory investigations of resistive-anode structure

Figure 8 shows the gain curves measured in bare- (metal) and resistive-WELL (RWELL) configurations (with no underlying Cu grid), compared to that of a regular THGEM with 2.3 mm induction gap and a 0.5 kV/cm induction field. The measurements were done with 8 keV x-rays in Ne/5%CH$_4$ at 1 atm. Note that there is no significant difference in gain between the bare- and resistive-WELL for the same voltage; the gain for a given voltage across the WELL is about 10-fold higher than in the regular THGEM. The maximum achievable gain for 8 keV x-rays (loosely defined as the gain at which a significant spark rate appears), is 1×10$^4$ for the regular THGEM, 2×10$^4$ for the bare WELL and 4×10$^4$ for the RWELL. This corresponds to a maximum achievable charge of 1.6×10$^6$, 3.2×10$^6$, and 6.4×10$^6$ electrons (about 250, 500 and 1000 fC) respectively. The higher gain for a given voltage in the WELL configuration, compared to the regular THGEM with an induction gap, results from the closed geometry which 'pushes' the field lines into the hole, yielding a larger field (and thus higher multiplication) near the anode (see [24] for more details).

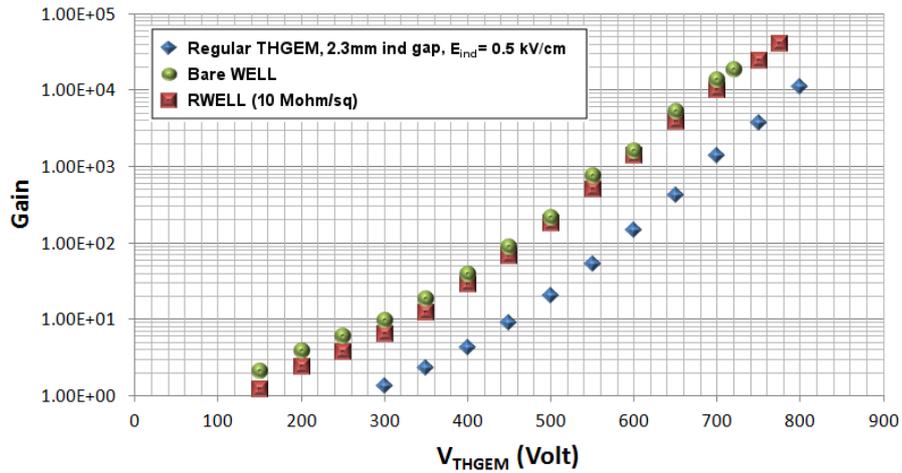

**Figure 8**: Gain curves for bare-WELL, resistive-WELL and regular THGEM with a 2.3 mm induction gap. In all cases the electrode parameters were: thickness 0.4 mm, hole diameter 0.5 mm, hole spacing 1 mm, rim size 0.1 mm. The gains were measured in Ne/5%CH$_4$ with 8 keV x-rays.

The measured discharge probabilities for the same three structures are displayed in figure 9. The onset of sparks occurred at a higher gain in the bare WELL compared to the normal THGEM with an induction gap and at still higher gains for the resistive WELL; for a given gain, the discharge probability was lower for the bare WELL compared to the normal THGEM and still lower for the RWELL structure. Note that for the RWELL (with no Cu grid) the average spark charge was ~15 fold lower than for the bare WELL (the charge was estimated by integrating the current provided by the HV power supply to the THGEM top electrode, following a discharge).

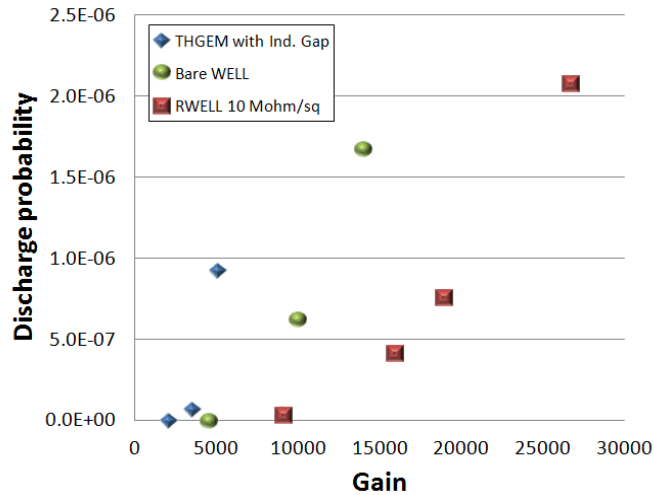

**Figure 9**: Discharge probability for the bare-WELL, resistive-WELL and regular THGEM with 2.3 mm induction gap (same geometry and conditions as in figure 8). The measurements were done with a 0.5 mm diameter collimated 8 keV x-ray beam, at a flux of $10^4$ Hz/mm$^2$.

Figure 10 shows the rate dependence of the gain for the RWELL structure. The gain drop observed is similar to that of previous measurements made with regular THGEM structures [25], suggesting that the main contribution to the loss of gain at high rates is the clearance time of the avalanche ions from the hole, which is of the order of 1 μs, and not the electron spread on the surface of the resistive layer. Note that in the range of expected rates in the ILC-DHCAL (<1 kHz/cm$^2$) the gain is essentially constant. Although the curve was measured for the RWELL (with no underlying Cu grid) the results of the gridded RWELL (GRWELL) can be expected to be rather similar.

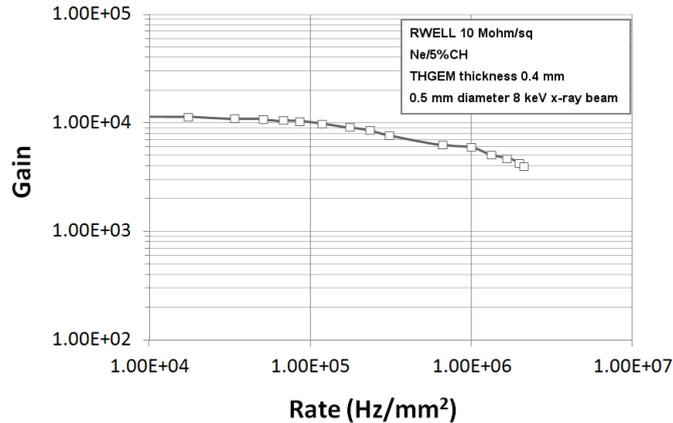

**Figure 10:** Rate dependence of the gain in the RWELL (10 MΩ/sq).

The cross talk between adjacent pads is displayed in figure 11a. Shown are cross talk data for the bare-WELL, resistive WELL (RWELL) and segmented gridded RWELL (segmented-GRWELL), as a function of the x-ray beam distance from the boundary between the pads. The measurements were performed by horizontally scanning the 0.5 mm diameter x-ray beam along a line connecting the pad centers. The pads were connected through charge sensitive preamplifiers and linear shaping amplifiers to a digital oscilloscope, with the amplifier gains balanced to yield similar average amplitudes when irradiating the center of both pads. The signal of the irradiated pad ('primary') was used as a trigger, and the signal of the neighboring pad, averaged over many pulses, was recorded. The figure shows the average neighbor pad signal normalized by that of the primary pad when the beam is at the primary pad's center. The figure demonstrates that when using an RWELL (with no Cu grid), the neighbor pad signal is larger than 10% of the primary signal up to a 3 mm distance from the pads boundary (which would result in large pad multiplicity). For a bare-WELL the neighbor pad signal is a few percent of that of the primary pad when the beam distance from the pads boundary is < 1 mm. For the segmented-GRWELL, there is essentially no observable cross talk between pads.

The same procedure also yielded information on the amplitude of the primary signal as a function of the beam position, as shown in figure 11b. For the bare-WELL the average amplitude of the primary signal is essentially constant up to a distance of ~1.5 mm from the pad boundary and then falls gradually to 50% exactly on it, reflecting symmetric charge sharing between pads. For the RWELL the signal gradually decreases over a ~1.5 mm distance from the boundary to 80% on it. For the segmented-GRWELL the signal begins to drop gradually at ~2.5-3 mm from the boundary, reaching 47% on it. In this case, the signal is somewhat smaller than 50% on the boundary, because not all of the primary electrons are collected into the holes on the opposite sides of the full copper strip above the pad boundary.

More details on the development of THGEM detectors with resistive anodes are given elsewhere [24].

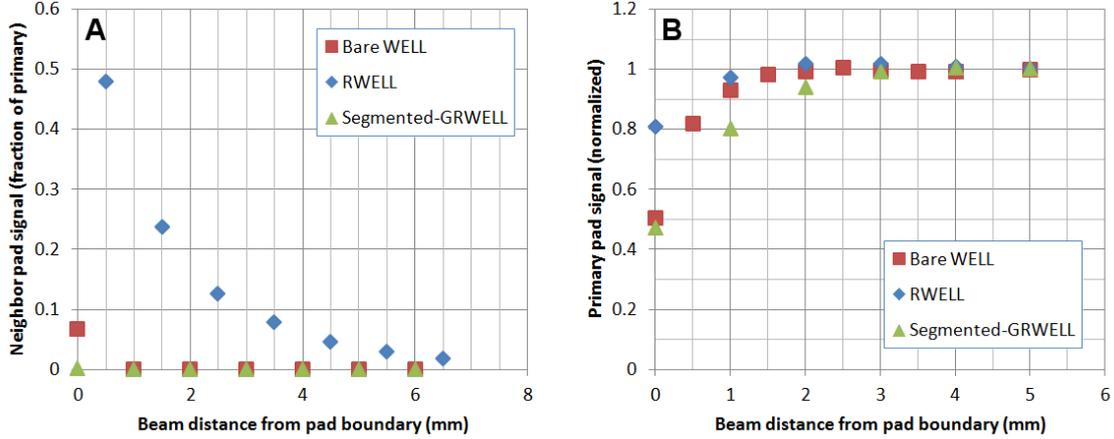

**Figure 11**: Neighbor pad (A) and primary pad (B) signal amplitude as a function of the beam distance from the boundary between two adjacent pads, for a bare WELL, resistive WELL (RWELL) and segmented gridded-RWELL. The beam (0.5 mm diameter 8 keV x-ray) was scanned across a line connecting the centers of two neighboring pads and the average signal amplitudes were recorded from both pads. The amplitudes are shown normalized by the value of the primary pad signal when the beam is exactly above its center.

## 4. Discussion

The results of the beam and laboratory studies demonstrate the potential of THGEM-based active elements for the ILC DHCAL, with competitive performances, simplicity and robustness.

The beam studies performed with low-noise discrete electronics and with an effective external trigger showed that a single-THGEM structure operating at a low gain ($\sim 1 \times 10^3$) can run at 96% efficiency with low discharge probability (of the order of $10^{-6}$ or lower). The particular configuration tested, which can still be optimized, had a total thickness of 5.5 mm excluding the readout electronics, well within the requirement of the ILC-DHCAL.

It was further demonstrated that a THGEM-based detector can be effectively operated with the KPiX readout system. Initial beam studies incorporating double-THGEM structures with gains in the range of $4\text{-}8\times 10^3$ showed stable operation with a discharge probability $< 2\times 10^{-6}$ for both muons and pions. The thinnest double-THGEM configuration tested (with a total width of 8 mm excluding readout electronics) can be further optimized, reducing its width to approach the DHCAL requirements. In this configuration the Landau MPV was at 14 fC for both muons and pions, indicating that very high efficiencies (close to unity) can be expected in the absence of empty triggers. Since KPiX noise is of the order of 0.3 fC, and because double-THGEM detectors are known to be more stable than single-THGEM ones, it can be safely estimated that such a detector can be operated with gains of the order of $1\text{-}2\times 10^3$, with discharge probabilities much lower than $10^{-6}$.

Beam studies done with a single-THGEM coupled to KPiX (with a total width of 6.5 mm excluding readout electronics), yielded Landau MPVs of 9 fC for both muons and pions. The discharge probability with the gain employed ($\sim 3\times 10^3$) was of the order of $\sim 1\times 10^{-4}$ for pions and $\sim 1\times 10^{-5}$ for muons – likely too high for the envisaged application. Reducing the gain to $\sim 1\times 10^3$ should enable a much more stable operation, similar to that obtained with a single-THGEM coupled to the discrete electronics. This would again require effective elimination of empty trigger events.

The pad multiplicity observed for the single-THGEM detector was ~1.1-1.2 for both muons and pions, for a threshold of 1.5 fC for the main hit and 1 fC for its neighbors. This level of multiplicity is similar to that obtained with MICROMEGAS-based active elements (1.1 [4]) and lower than that observed for RPCs (~1.5-2 [3]). Similar multiplicity values are foreseen for GEM-based active elements [5]. The 96% efficiency obtained with a single-THGEM with discrete readout is close to that obtained with MICROMEGAS (98% [4]) and somewhat larger than that of RPCs (~92% [3]), where the MICROMEGAS and RPC efficiencies correspond to the multiplicities given above. With a threshold at 1-1.5 fC and MPV at 2.5-5 fC (gain of $\sim 1$-$2\times 10^3$), similar efficiencies (~95%) are anticipated for a single-THGEM with KPiX, provided empty triggers are eliminated. The interplay between multiplicity and efficiency will be further studied in future beam tests.

Efforts are presently undertaken to optimize the geometry of the double-THGEM configuration, allowing for highly stable operation with a gain that would provide near unity efficiencies, at reduced thickness. These include, along with a reduction of the electrode spacing, an investigation of the optimal operation parameters with Ar-based mixtures, for which the total number of deposited electrons per unit length is ~ 2.4 fold larger compared to Ne-based mixtures [12]. (One should note, however, that with Ar-based mixtures the charge deposited by highly ionizing particles will also be correspondingly larger).

The R&D efforts on WELL type detectors, with either bare metal pads or resistive anode structures [24], may lead to very thin detectors configurations (4-5 mm excluding readout). The bare WELL structure was shown to have a lower discharge rate for a given gain, compared to the regular single-THGEM with an induction gap. Such a detector can be expected to have similar efficiencies and pad multiplicity as those demonstrated with a regular single-THGEM detector. Its only potential drawback, compared to the regular single-THGEM with an induction gap, is that discharge-induced energy is fully directed towards the readout pads, while with the induction gap, a large fraction of it is collected by the bottom-THGEM electrode [24]. This may be mitigated by improving the spark protection of the readout board.

Studies on resistive-anode structures [24] aim at protecting the electronics from occasional discharge-induced transients, allowing, as shown here, for detector operation at appropriate gains with significantly reduced discharge energies and probabilities. Of the resistive anode configurations investigated so far, the segmented gridded resistive WELL (segmented-GRWELL) appears to be the most promising. Like the bare WELL, this configuration may allow for conceiving very thin, stable active elements. The present challenges are to demonstrate its good efficiency and multiplicity scores with suitable readout electronics. The challenge would be keeping high detection efficiency close to pad boundaries where losses of induced-signal amplitude occur. Optimization of the segmented-GRWELL parameters is ongoing both experimentally and by Monte Carlo simulations.

**Acknowledgments**

This work was supported in part by the Israel-USA Binational Science Foundation (Grant 2008246), by the Benozyio Foundation and by the FCT Projects PTDC/FIS/113005/2009 and CERN/FP/116394/2010. The research was done within the CERN RD51 collaboration. H. Natal da Luz is supported by FCT grant SFRH/BPD/66737/2009. C. D. R Azevedo is supported by the SFRH/BD/35979/2007 grant and D. S. Covita by the SFRH/BPD/46611/2008 grant through FCT and FEDER programs. A. Breskin is the W.P. Reuther Professor of Research in the Peaceful use of Atomic Energy. The THGEM electrodes were manufactured at the CERN Printed Circuits Workshops; we are indebted to Mr. Rui de Oliveira for his continuous invaluable cooperation. We also wish to thank CERN's Gas Detectors Development group for its support during the beam tests.